\documentclass[12pt,preprint]{aastex}

\begin{document}
\title{The $\Lambda$CDM model on the lead -- a Bayesian cosmological models comparison}
\author{Aleksandra Kurek}
\affil{Astronomical Observatory, Jagiellonian University, Orla 171, 30-244 Krak{\'o}w, Poland}

\author{Marek Szyd{\l}owski}
\affil{Astronomical Observatory, Jagiellonian University, Orla 171, 30-244 Krak{\'o}w, Poland}
\affil{Complex Systems Research Centre, Jagiellonian University, Reymonta 4, 30-059 Krak{\'o}w, Poland}

\begin{abstract}
Recent astronomical observations indicate that our Universe is undergoing a period of an accelerated expansion. While there are many cosmological models, which explain this phenomenon, the main question remains which is the best one in the light of available data. We consider ten cosmological models of the accelerating Universe and select the best one using the Bayesian model comparison method. We demonstrate that the $\Lambda$CDM model is most favored by the Bayesian statistical analysis of the SNIa, CMB, BAO and H(z) data.
\end{abstract}

\section{Introduction}

Recent observations of type Ia a supernovae (SNIa) provide the main evidence that the current Universe is in an accelerating phase of expansion \citep{Riess:1998cb, Perlmutter:1998np}. There are many different cosmological models used in explanation of an accelerating phase of evolution of the current Universe. They can be divided into two groups of models according to `philosophical' assumptions on a cause of the accelerated expansion. In the first type of explanation the conception of mysterious dark energy of an unknown form is used, while in the second one it is postulated some modification of the Friedmann equation. Here we choose five models which belong to the former group as well as five ones which belong to the latter one. All the chosen models are assumed to be spatially flat.

If we assume the Friedmann-Robertson-Walker (FRW) model in which effects of non-homogeneities are neglected, than acceleration can be driven by a dark energy component $X$ (matter fluid violating the strong energy condition). This kind of energy represents roughly $70\%$ of the matter content of the present Universe. The model with the cosmological constant (the $\Lambda$CDM model) has the equation of state for dark energy as follows: $p_X=-\rho_X$ \citep{Weinberg:1989}. The model with phantom dark energy has $p_X=w_X\rho_X$, where $w_X$ ($< -1$) is a negative constant \citep{Caldwell, Dabrowski}. The next one is the model with a dynamical coefficient of the equation of state, parameterized by the scale factor $a$: $w(a)=w_0+w_1(1-a)$ \citep{Polarski2001, Linder2003}. The other simple approach is to represent dark energy in the form of a minimally coupled scalar field $\phi$ with the potential $V(\phi)$. In cosmology the quintessence idea is important in understanding a role of the scalar field in the current Universe. We consider the power-law parameterized quintessence model \citep{Peebles:1988, Ratra:1988}. In this case density of dark energy changes with the scale factor as $\rho_X=\rho_{X0}a^{-3(1+\bar{w}_{X}(a))}$, where $\bar{w}_{X}(a)$ is the mean of a coefficient of the equation of state in the logarithmic scale factor
\begin{displaymath}
\bar{w}_X(a)=\frac{\int w_X(a)d\ln a}{\int d(\ln a)}
\end{displaymath}
and has the following form $\bar{w}_X = w_0 a^{\alpha}$ \citep{Rahvar:2007}.
The first group is completed with the model with the generalized Chaplygin gas, where $p_X=-\frac{A}{\rho_X^{\alpha}}$ (here $A>0$ and $\alpha=\mathrm{const}$).
We gathered above models together with their Hubble functions (with the assumption that the Universe is spatially flat) in Table~\ref{tab:1}.

\begin{deluxetable}{rll}
\tabletypesize{\scriptsize}
\tablecaption{The Hubble function for cosmological models with dark energy\label{tab:1}}
\tablewidth{0pt}
\tablehead{\colhead{case}&\colhead{model}&\colhead{$H^{2}(z)$relation}} 
\startdata
1 & $\Lambda$CDM model&
 $H^{2}=H^{2}_{0}\{\Omega_{\mathrm{m},0}(1+z)^{3}+(1- \Omega_{\mathrm{m},0})\}$\\ 
  & & \\
2&model with generalized Chaplygin gas&
 $H^{2}=H^{2}_{0} \left\{\Omega_{\mathrm{m},0}(1+z)^{3} + (1-\Omega_{\mathrm{m},0})[A_{S}+(1-A_{S})
 (1+z)^{3(1+\alpha)}]^{\frac{1}{1+\alpha}}\right\}$ \\
 & & \\
3&model with phantom dark energy & 
 $H^{2}=H^{2}_{0}\{\Omega_{\mathrm{m},0}(1+z)^{3}+ 
 (1-\Omega_{\mathrm{m},0})(1+z)^{3(1+w_{X})}\}$ \\
 & & \\
4&model with dynamical E.Q.S&
 $H^{2}=H^{2}_{0} \left\{\Omega_{\mathrm{m},0}(1+z)^{3}+ 
 (1-\Omega_{\mathrm{m},0})(1+z)^{3(w_{0}+w_{1}+1)} \exp [- \frac {3w_{1}z}{1+z}] \right\}$ \\
 & & \\
5&quintessence model& 
 $H^{2}=H^{2}_{0} \left\{\Omega_{\mathrm{m},0}(1+z)^{3}+ 
 (1-\Omega_{\mathrm{m},0})(1+z)^{3(1+w_0(1+z)^{-\alpha})} \right\}$ \\
\enddata
\end{deluxetable}

As we have written before we also consider five models offering the explanation of current acceleration of the Universe in an alternative way to dark energy. The brane models has postulated that the observer is embedded on the brane in a larger space in which gravity can propagate: the Dvali-Gabadadze-Porrati model (DGP) \citep{Dvali:2000}, Sahni-Shtanov brane 1 model \citep{Shtanov:2000}. The Cardassian model, in which the Universe is flat, is matter dominated and accelerating as a consequence of the modification of the Friedmann first integral as follows $3H^2=\rho+B \rho ^n$, where $B$ is a constant and the energy density contains only dust matter and radiation \citep{Freese:2002}. We also include in the analysis the bouncing model arising in the context of loop quantum gravity (the B$\Lambda$CDM model) \citep{Singh:2005, Szydlowski:2005} and the model with energy transfer between the dark matter and dark energy sectors (the $\Lambda$ decaying vacuum model) \citep{Szydlowski:2006}. We gathered above models together with their Hubble functions (with the assumption that the Universe is spatially flat) in Table~\ref{tab:2}.

\begin{deluxetable}{lll}
\tabletypesize{\scriptsize}
\tablecaption{The Hubble function for cosmological models with modified theory of gravity\label{tab:2}}
\tablewidth{0pt}
\tablehead{\colhead{case}&\colhead{model}&\colhead{$H^{2}(z)$relation}} 
\startdata
6& DGP model &
 $H^{2}=H^{2}_{0} \left\{ \left [ \sqrt{\Omega_{\mathrm{m},0}(1+z)^{3}+\Omega_{rc,0}}+
 \sqrt{\Omega_{rc,0}} \right] ^{2} \right\}$ \\    
 & & $\Omega_{rc,0}=\frac{(1-\Omega_{\mathrm{m},0})^2}{4}$ \\
 & & \\
7& B$\Lambda$CDM model &
 $H^{2}=H^{2}_{0} \left\{ \Omega_{\mathrm{m},0}(1+z)^{3}- 
 \Omega_{n,0}(1+z)^{n}+1-\Omega_{\mathrm{m},0}+\Omega_{n,0} \right\}$ \\
 & & \\
8& interacting model with $\Lambda$&
  $H^{2}=H^{2}_{0} \{ \Omega_{\mathrm{m},0}(1+z)^{3}+\Omega_{\mathrm{int},0}(1+z)^{n}+ 
  1-\Omega_{\mathrm{m},0}-\Omega_{\mathrm{int},0}\}$ \\
 & & \\
9& Cardassian model&
 $H^{2}=H^{2}_{0} \left\{ \Omega_{r,0}(1+z)^{4}+\Omega_{\mathrm{m},0}(1+z)^{4} 
 \left[ \frac{1}{1+z} + (1+z)^{-4+4n} \left ( \frac{1- \Omega_{r,0}-
 \Omega_{\mathrm{m},0}}{\Omega_{\mathrm{m},0}} \right ) \left ( \frac{ \frac{1}{1+z}+
 \frac{\Omega_{r,0}}{\Omega_{\mathrm{m},0} }}{ 1+ 
 \frac{\Omega_{r,0}}{\Omega_{\mathrm{m},0}}}\right)^{n} \right ] \right\}$ \\
 & $\Omega_{r,0}=10^{-4}$ & \\
 & & \\
10& Sahni-Shtanov brane I model&
 $H^{2}=H^{2}_{0} \left\{ \Omega_{\mathrm{m},0}(1+z)^{3}+\Omega_{\sigma,0}+ 
 2 \Omega_{l,0}-2\sqrt{\Omega_{l,0}}\sqrt{\Omega_{\mathrm{m},0}(1+z)^{3}+
 \Omega_{\sigma,0}+ \Omega_{l,0}+ \Omega_{\Lambda b,0}} \right\}$ \\
& & \\
&& $\Omega_{\sigma,0}=1-\Omega_{\mathrm{m},0}+2\sqrt{\Omega_{l,0}}\sqrt{1+\Omega_{\Lambda b,0}}$ \\
\enddata
\end{deluxetable}

The main goal of this paper is to compare all these models in the light of SNIa, CMB, BAO and H(z) data. We use the Bayesian model comparison method, which we describe in the next section. This method is commonly used in the context of cosmological models selection \citep[see e.g.][]{Liddle:2004, John:2002gg, Saini:2003wq, Parkinson:2004yx, Mukherjee:2005tr, Beltran:2005xd, Mukherjee:2005wg, Szydlowski:2005xv, Godlowski:2005tw, Szydlowski:2006xx, Liddle:2006, Liddle:2007, Sahlen:2007, Serra:2007id, Kunz:2006, Trotta:2007, Trotta:2007a}.
Recently the Bayesian information criteria were applied in the context of choosing an adequate model of acceleration of the Universe \citep{Davis:2007na}. The authors showed preference for models beyond the standard FRW cosmology (so-called exotic cosmological models) whose best fit parameters reduce them to the cosmological constant model.

\section{Model comparison in Bayes theory}

Let us consider the set of $K$ models: $\{ M_1, \cdots, M_K \}$. In the Bayes theory the best model from the set under consideration is this one which has the largest value of the probability in the light of the data ($D$), so called posterior probability
\begin{equation}\label{eq:5}
P(M_i|D)=\frac{P(D|M_i)P(M_i)}{P(D)}.
\end{equation}
$P(M_i)$ is the prior probability for the model indexed by $i$, which value depends on our previous knowledge about model under consideration, that is to say without information coming from data $D$, and $P(D)$ is the normalization constant. If we have no foundation to favor one model over another one from the set we usually assume the same values of this quantity for all of them, i.e. $P(M_i)=\frac{1}{K},\ i=1,\cdots,K$.

To obtain the form of $P(D)$ it is required that a sum of the posterior probabilities for all models from the set is equal one
\begin{displaymath}
\sum _{i=1}^{K} P(M_i|D)=1 \  \longrightarrow P(D)=\sum_{i=1}^{K}P(D|M_i)P(M_i).
\end{displaymath}
Therefore conclusions based on the values of posterior probabilities strongly depend on the set of models and can change when the set of models is different.
 
$P(D|M_i)$ is the marginal likelihood (also called the evidence) and has the following form
\begin{equation}\label{eq:1}
P(D|M_i)=\int L(\bar{\theta _i}|D,M_i)P(\bar{\theta_i}|M_i)d\bar{\theta_i} \equiv E_{i},
\end{equation}
where $L(\bar{\theta _i}|D,M_i)$ is the likelihood of the model under consideration, $\bar{\theta _i}$ is the vector of the model parameters and $P(\bar{\theta_i}|M_i)$ is the prior probability for the model parameters.

In the case under consideration we cannot obtain the value of the evidence by analytical computation. We need a numerical method or an approximation to this quantity.

Schwarz \citet{Schwarz:1978} showed that for iid observations ($D=\{x_i\}, i=1,\cdots,N$) coming from a linear exponential family distribution, defined as
\begin{displaymath}
 f(x_{i}|\bar{\theta})= \exp \left [\sum _{k=1}^{S}w_{k}(\bar{\theta})t_{k}(x_{i})+b(\bar{\theta})\right ], \quad S=d,
\end{displaymath}
where $w_{1},\dots,w_{S},b$ are functions of only $\bar{\theta}\in \mathbf{R}^{d}$,  $t_{1},\dots,t_{S}$ are function of only $x_{i}$, the asymptotic approximation ($N\to \infty$) to the logarithm of the evidence is given by
\begin{equation}\label{eq:2}
\ln E = \ln \mathcal{L} - \frac{d}{2}\ln N + O(1),
\end{equation} 
where $\mathcal{L}$ is the maximum likelihood and $O(1)$ is the term of order unity in $N$. In this case the likelihood function has the following form 
\begin{displaymath}
L(\bar{\theta}|D,M)=\Pi_{i=1}^{N} f(x_{i}|\bar{\theta})=\exp \left[ N \left( \sum_{k=1}^{S}w_{k}(\bar{\theta})T_{k}(D)+b(\bar{\theta})\right)\right],
\end{displaymath}
where $T_{k}(D)=\frac{1}{N}\sum_{i=1}^{N}t_{k}(x_{i})$. The integral (\ref{eq:1}) can be writing as 
\begin{equation}\label{eq:3}
\int \exp[ N g(\bar{\theta})]P(\bar{\theta}|M) d \bar{\theta},
\end{equation}
where $g(\bar{\theta})=\sum_{k=1}^{S}w_{k}(\bar{\theta})T_{k}(D)+b(\bar{\theta})$.
This integral has the form of the so called Laplace integral. Assume that $g(\bar{\theta})$ has maximum in $\bar{\theta}_{0}$ and $P(\bar{\theta}_{0}|M)\ne 0$. When $N \to \infty$  $\exp[Ng(\bar{\theta})]$ will be a sharp function picked at $\bar{\theta}_{0}$. Then the main contribution to integral (\ref{eq:3}) comes from the small neighborhood of $\bar{\theta}_{0}$. In this region  $P(\bar{\theta}|M)\approx P(\bar{\theta}_{0}|M)$, we can also replace $g(\bar{\theta})$ function its Taylor expansion around $\bar{\theta}_{0}$: $g(\bar{\theta})= g(\bar{\theta}_{0})-\frac{1}{2}(\bar{\theta}-\bar{\theta}_{0})^{T}C^{-1}(\bar{\theta}-\bar{\theta}_{0})$, where $ \left[C^{-1}\right]_{ij}=\left[-\frac{\partial^{2} g(\bar{\theta})}{\partial \theta _{i} \partial \theta_{j}}\right]_{\bar{\theta}=\bar{\theta}_{0}}$ and extend the integration region to whole $\mathbf{R}^{d}$. One can gets the asymptotic of the integral (\ref{eq:3}) $E=\exp[Ng(\bar{\theta}_0)]P(\bar{\theta}_{0}|M)\left(\frac{2\pi}{N}\right)^{\frac{d}{2}}\sqrt{detC}$ and $\ln E= Ng(\bar{\theta}_{0})-\frac{d}{2}\ln N + R $, where $R$ is the term which not depend on N. One can see that $ Ng(\bar{\theta}_0)=\ln L(\bar{\theta}_{0}|D,M)$, where $\bar{\theta}_{0}$ is the point which maximize $g(\bar{\theta})=\frac{1}{N}\ln L(\bar{\theta}|D,M)$, so is equivalent to $\bar{\theta}_{MLE}$ (the maximum likelihood estimator of $\bar{\theta}$). Finally one can obtain result (\ref{eq:2}).

According to this result Schwarz introduced a criterion for the model selection: the best model is that which minimizes the $BIC$ quantity, defined as
\begin{equation}
BIC=-2 \ln \mathcal{L} + d \ln N.
\end{equation}

This criterion can be derived in such a way that it is not required to assume any specific form for the likelihood function but it is only necessary that the likelihood function satisfies some non-restrictive regularity conditions. Moreover data do not need to be independent and identically distributed. This derivation requires to assume that a prior for model parameters is not equal to zero in the neighborhood of the point where the likelihood function under a given model reaches a maximum and that it is bound in the whole parameter space under consideration \citep{Cavanaugh:1999}. It should be pointed out that an asymptotic assumption is satisfied when a sample size used in analysis is large with respect to the number of unknown model parameters.

It is useful to choose one model from our models set (here indexed by $s$) and compare the rest models with this one. We can define $\Delta BIC_{is}$ quantity, which is the difference of the $BIC$ quantity for the models indexed by $i$ and $s$: $\Delta BIC_{is}=BIC_i -BIC_s$ and present the posterior probability in the following form
\begin{equation}\label{eq:4}
P(M_i|D)=\frac{\exp(-\frac{1}{2}\Delta BIC_{is})P(M_i)}{\sum _{k=1}^K \exp(-\frac{1}{2}\Delta BIC_{ks})P(M_k)}.
\end{equation}

Let us assume that we have computed the probabilities in the light of data $D$ for models from the set under consideration. Then we gathered new data $D_1$ and want to update the probabilities which we already have. We can compute probabilities in the light of new data using information coming from previous analysis, which allow us to favor one model over another: we can use posterior probabilities for models obtained in earlier computations as a prior probabilities for models in next analysis.

We apply this method in evaluation the posterior probabilities for models described in the previous section using the information coming from SNIa, CMB, BAO and $H(z)$ data.

\section{Application to cosmological models comparison}

We start with the $N=192$ sample of SNIa \citep{Riess:2007, Wood:2007, Davis:2007}. In this case the likelihood function has the following form
\begin{displaymath}
 L\propto \exp \left[-\frac{1}{2}\left(\sum_{i=1}^{N}\frac{(\mu_{i}^{\mathrm{theor}}-\mu_{i}^{obs})^{2}}{\sigma_{i}^{2}}\right) \right],
\end{displaymath}
where $\sigma_{i}$ is known, $\mu_{i}^{obs}=m_{i}-M$ ($m_{i}$--the apparent magnitude, $M$--the absolute magnitude of SNIa), $\mu_{i}^{\mathrm{theor}}=5\log_{10}D_{Li} + \mathcal{M}$, $\mathcal{M}=-5log_{10}H_{0}+25$ and $D_{Li}=H_{0}d_{Li}$, where $d_{Li}$ is the luminosity distance, which with assumption that $k=0$ is given by
\begin{displaymath}
d_{Li}=(1+z_{i})c\int_{0}^{z_{i}} \frac{d z'}{H(z')}.
\end{displaymath}
In this case we used the BIC quantity as an approximation to the minus twice logarithm of evidence and assumed that all models have equal values of prior probabilities. Based on our previous experiences we have used following assumptions for models parameters values: $H_{0} \in \langle 60, 80 \rangle$ for all models and additional:
\begin{itemize}
    \item model 2: $A_{S} \in \langle 0,1 \rangle$, $\alpha \in \langle 0,1 \rangle$ 
    \item model 3: $w_{X} \in \langle -4,-1 )$
    \item model 4: $w_{0} \in \langle -3,3 \rangle$, $w_{1} \in \langle -3,3 \rangle$
    \item model 5: $w_{0} \in \langle -3,3 \rangle$, $\alpha \in \langle 0,2 
\rangle$
    \item model 7: $\Omega_{n,0} \in \langle 0,1 \rangle$, $n \in \langle 3,10 
\rangle$  
    \item model 8: $\Omega_{\mathrm{int},0} \in \langle -1,1 \rangle$, $n \in \langle -10,10 \rangle$ 
    \item model 9: $n \in \langle -10,10 \rangle$
    \item model 10: $\Omega_{l,0} \in \langle 0,4 \rangle$, $\Omega_{\Lambda b,0} \in \langle -1,4 \rangle$  
\end{itemize}
We separately consider cases with $\Omega_ {\mathrm{m},0} \in \langle 0,1 \rangle$ and  
$\Omega_ {\mathrm{m},0} \in \langle 0.25,0.31 \rangle$. We treat the $H_0$ parameter as a nuisance parameter, i.e. we marginalized the likelihood function over this parameter in the range assumed before. Posterior probabilities are obtained using equation \ref{eq:4}. We analyse three sets of models: 1. set of models with dark energy (Table~\ref{tab:1}), 2. set of models with modified theory of gravity (Table~\ref{tab:2}), 3. set of all models (Table~\ref{tab:1} and Table~\ref{tab:2} together).

The results for the case with $\Omega_{\mathrm{m},0} \in \langle 0, 1 \rangle $ are presented in Table 3, Table 4 and Table 5 for set 1, set 2 and set 3 respectively. One can conclude that in the light of SNIa data the $\Lambda$CDM is the best model from the set of models with dark energy as well as the best one from the all models under consideration, the DGP model is the best one from the group of models with modified gravity.

The results for case with  $\Omega_{\mathrm{m},0} \in \langle 0.25, 0.31 \rangle $ are gathered in Table 6, Table 7, Table 8 for set 1, set 2 and set 3 respectively. The conclusion changed for the set of models with modified gravity: here the best one is the Cardassian model.

In the next step we included information coming from CMB data. Here the likelihood function has the following form
\begin{displaymath}
L \propto \exp \left[- \frac {(R^{\mathrm{theor}}-R^{\mathrm{obs}})^2}{2\sigma_{R}^2} \right],
\end{displaymath}
where $R$ is so called shift parameter, $R^{\mathrm{theor}}=\sqrt{\Omega_{\mathrm{m},0}}\int_{0}^{z_{dec}}\frac{H_0}{H(z)}dz$, and $R^{\mathrm{obs}}=1.70 \pm 0.03$ for $z_{\mathrm{dec}}=1089$ \cite{Spergel:2006, Wang:2006ts}. It should be pointed out that the parameter $R$ is independent of $H_{0}$.

The values of the evidence were obtained by the numerical integration. We assumed flat prior for all model parameters. It is known that evidence depends on the prior probabilities for model parameters. Assumptions for the model parameters intervals, which we made in previous analysis could be not appropriate here. Due to this we made a stricter analysis for models with parameters which interval width exceeds one. This width was used for convenience. We computed the evidence for different parameter intervals, which do not exceed the range assumed before and with a minimal width equal to one. There are of course extremely many possibilities. We limited our analysis to intervals $\langle a,b \rangle$, where $a$ and $b$ are integer. Finally we chose the case with the greatest evidence.

We consider the situation with $\Omega_ {\mathrm{m},0} \in \langle 0,1 \rangle$ and  
$\Omega_ {\mathrm{m},0} \in \langle 0.25,0.31 \rangle$. The range for parameters which change after stricter analysis in the first case:
\begin{itemize}
    \item model 3: $w_{X} \in \langle -2,-1 )$
    \item model 4: $w_{0} \in \langle -1,0 \rangle$, $w_{1} \in \langle -2,0 \rangle$
    \item model 5: $w_{0} \in \langle -3,-2 \rangle$, $\alpha \in \langle 1,2 
\rangle$
    \item model 7: $\Omega_{n,0} \in \langle 0,1 \rangle$, $n \in \langle 3,4 
\rangle$  
    \item model 8: $\Omega_{\mathrm{int},0} \in \langle -1,0 \rangle$, $n \in \langle -10,-9 \rangle$ 
    \item model 9: $n \in \langle 0,1 \rangle$
    \item model 10: $\Omega_{l,0} \in \langle 0,1 \rangle$, $\Omega_{\Lambda b,0} \in \langle 0,1 \rangle$  
\end{itemize}
and in the second case:
\begin{itemize}
    \item model 3: $w_{X} \in \langle -2,-1 )$
    \item model 4: $w_{0} \in \langle -1,0 \rangle$, $w_{1} \in \langle -2,-1 \rangle$
    \item model 5: $w_{0} \in \langle -2,-1 \rangle$, $\alpha \in \langle 0,1 
\rangle$
    \item model 7: $\Omega_{n,0} \in \langle 0,1 \rangle$, $n \in \langle 3,4 
\rangle$  
    \item model 8: $\Omega_{\mathrm{int},0} \in \langle -1,0 \rangle$, $n \in \langle -2,-1 \rangle$ 
    \item model 9: $n \in \langle 0,1 \rangle$
    \item model 10: $\Omega_{l,0} \in \langle 0,1 \rangle$, $\Omega_{\Lambda b,0} \in \langle 3,4 \rangle$  
\end{itemize}

Posterior probabilities were obtained using equation \ref{eq:5}. Here we treated posterior probabilities evaluated in analysis with SNIa data as a prior probabilities. Results are gathered in tables like in previous analysis. We also show the values of the posterior probabilities obtained for the intervals assumed at the beginning (numbers in the brackets). As we can see the $\Lambda$CDM model is still the best one from the models with dark energy (for both ranges of $\Omega_{\mathrm{m},0}$). The conclusion is the same for the set of models with modified gravity: the DGP model is the best one in the first case and the Cardassian model in the second. When we assume that $\Omega_ {\mathrm{m},0} \in \langle 0,1 \rangle$ there is no evidence to favor the $\Lambda$CDM model over the DGP model (they have the same values of the posterior probabilities) while when we restrict $\Omega_ {\mathrm{m},0}$ range to $\langle 0.25,0.31 \rangle$ the $\Lambda$CDM model still stays as the best one, with even greater probability.

As the third observational data we used the measurement of the baryon acoustic oscillations (BAO) from the SDSS luminous red galaxies \citep{Eisenstein:2005}. In this case the likelihood function has the following form
\begin{displaymath}
L \propto \exp \left[ -\frac{(A^{\mathrm{theor}}-A^{\mathrm{obs}})^2}{2\sigma_{A}^2}\right ],
\end{displaymath}
where $A^{\mathrm{theor}}=\sqrt{\Omega_{\mathrm{m},0}} \left (\frac{H(z)}{H_{0}} \right ) ^{-\frac{1}{3}} \left [ \frac{1}{z_{A}} \int_{0}^{z_{A}}\frac {H_0}{H(z)}\right]^{\frac{2}{3}}$ and $A^{obs}=0.469 \pm 0.017$ for $z_{A}=0.35$.

Here values of the evidence are obtained by the numerical integration.
We made the analogous analysis with the parameters intervals as above with the additional requirement: obtained intervals must at least cover the intervals obtained in previous analysis. For most of models the conclusions are the same as in the previous analysis. Below we wrote the cases where the intervals have changed \\
for the case with $\Omega_ {\mathrm{m},0} \in \langle 0,1 \rangle$: 
\begin{itemize}
\item model 5: $w_{0} \in \langle -3,1 \rangle$, $\alpha \in \langle 1,2 
\rangle$
\end{itemize}
and with $\Omega_ {\mathrm{m},0} \in \langle 0.25,0.31 \rangle$:
\begin{itemize}
    \item model 8: $\Omega_{\mathrm{int},0} \in \langle -1,0 \rangle$, $n \in \langle -2,1 \rangle$  
\end{itemize}

Here we used posterior probabilities obtained in analysis with the CMB data as prior probabilities. Results were again presented in described above tables. The conclusion is different for the set of all models in the case with $\Omega_ {\mathrm{m},0} \in \langle 0,1 \rangle$: the DGP model becomes the best one from them.

Finally we used the observational $H(z)$ data ($N=9$) from Simon et al. \citet{Simon:2005} \citep[see also][and references therein]{Samushia:2006fx, Wei:2007}. These data based on the differential ages ($\frac{dt}{dz}$) of the passively evolving galaxies which allow to estimate the relation $H(z)\equiv \frac{\dot{a}}{a}=-\frac{1}{1+z}\frac{dz}{dt}$. Here the likelihood function has the following form
\begin{displaymath}
L \propto \exp \left(-\frac{1}{2} \left[ \sum_{i=1}^{N}\frac{\left( H(z_i) -H_i(z_i) \right) ^2}{\sigma_i ^2} \right] \right),
\end{displaymath}
where $H(z)$ is the Hubble function, $H_i,\ z_i$ are observational data.

The values of the evidence were obtained by the numerical integration. As above we assumed flat prior probabilities for models parameters. The ranges for them which changed after analogous to previous analysis are as follows:\\
for case with $\Omega_ {\mathrm{m},0} \in \langle 0,1 \rangle$ and $H_0 \in \langle 60,80 \rangle$:
\begin{itemize}
\item model 5: $w_{0} \in \langle -3,1 \rangle$, $\alpha \in \langle 0,2 
\rangle$
\item model 8: $\Omega_{\mathrm{int},0} \in \langle -1,0 \rangle$, $n \in \langle -10,0 \rangle$  
\end{itemize}
and for case with $\Omega_ {\mathrm{m},0} \in \langle 0.25,0.31 \rangle$ and $H_0 \in \langle 60,80 \rangle$:
\begin{itemize}
\item model 4: $w_{0} \in \langle -1,0 \rangle$, $w_{1} \in \langle -2,0 \rangle$  
\end{itemize}

Values of posterior probabilities obtained in analysis with the BAO data were treated as prior probabilities it this analysis.
The results are presented in tables.

As one can see in the case with $\Omega_ {\mathrm{m},0} \in \langle 0.25,0.31 \rangle$ the $\Lambda$CDM model is the best one from the set of models with dark energy as well as the best one from all models considered in this paper. The conclusion is different in the set of the models with modified gravity: after the analysis with observational $H(z)$ data the DGP model becomes the best one. In the case with  $\Omega_ {\mathrm{m},0} \in \langle 0,1 \rangle$ the $\Lambda$CDM is still the best model from the set of models with dark energy while the DGP model is the best one from the set of models with the modified theory of gravity as well as the best one from all models considered.

\begin{deluxetable}{cccccc|c}
\tabletypesize{\scriptsize}
\tablecaption{Posterior probabilities for models from Table~\ref{tab:1}; $\Omega_ {\mathrm{m},0} \in \langle 0,1 \rangle$ }
\tablewidth{0pt}
\tablehead{\colhead{model}&\colhead{prior}&\colhead{posterior SNIa}& \colhead{posterior + CMB}&\colhead{posterior + BAO}&\colhead{posterior + H(z)}&\colhead{posterior SNIa+CMB+BAO+H(z)}}
\startdata
1 & 0.20 & 0.91 & 0.92 (0.95) & 0.91 (0.97) & 0.94 (0.99)& 0.84\\
2 & 0.20 & 0.01 & 0.01 (0.01) & 0.01 (0.01) & 0.00 (0.00)& 0.02\\
3 & 0.20 & 0.07 & 0.06 (0.04) & 0.07 (0.03) & 0.05 (0.01) & 0.06\\
4 & 0.20 & 0.01 & 0.01 (0.00) & 0.01 (0.00) & 0.01 (0.00)& 0.04\\
5 & 0.20 & 0.00 & 0.00 (0.00) & 0.00 (0.00) & 0.00 (0.00) & 0.04\\
\enddata
\end{deluxetable}

\begin{deluxetable}{cccccc|c}
\tabletypesize{\scriptsize}
\tablecaption{Posterior probabilities for models from Table~\ref{tab:2}; $\Omega_ {\mathrm{m},0} \in \langle 0,1 \rangle$ }
\tablewidth{0pt}
\tablehead{\colhead{model}&\colhead{prior}&\colhead{posterior SNIa}& \colhead{posterior + CMB}&\colhead{posterior + BAO}&\colhead{posterior + H(z)}&\colhead{posterior SNIa+CMB+BAO+H(z)}}
\startdata
6 & 0.20 & 0.89 & 0.92 (0.97)& 0.92 (0.98)& 0.93 (0.98)& 0.07 \\
7 & 0.20 & 0.01 & 0.00 (0.00)& 0.00 (0.00)& 0.00 (0.00) & 0.03\\
8 & 0.20 & 0.01 & 0.01 (0.01)& 0.01 (0.01)& 0.01 (0.01) & 0.13\\
9 & 0.20 & 0.08 & 0.07 (0.01)& 0.07 (0.00)& 0.06 (0.00) & 0.74\\
10 & 0.20 & 0.01 & 0.00 (0.01)& 0.00 (0.01)& 0.00 (0.01) & 0.03\\
\enddata
\end{deluxetable}

\begin{deluxetable}{cccccc|c}
\tabletypesize{\scriptsize}
\tablecaption{Posterior probabilities for models from Table~\ref{tab:1} and Table~\ref{tab:2}; $\Omega_ {\mathrm{m},0} \in \langle 0,1 \rangle$ }
\tablewidth{0pt}
\tablehead{\colhead{model}&\colhead{prior}&\colhead{posterior SNIa}& \colhead{posterior + CMB}&\colhead{posterior + BAO}&\colhead{posterior + H(z)}&\colhead{posterior SNIa+CMB+BAO+H(z)}}
\startdata
1 & 0.10 & 0.51 & 0.46 (0.48)& 0.44 (0.46)& 0.43 (0.45) & 0.74\\
2 & 0.10 & 0.00 & 0.00 (0.00)& 0.00 (0.00)& 0.00 (0.00) &0.02\\
3 & 0.10 & 0.04 & 0.03 (0.02)& 0.03 (0.01)& 0.02 (0.00) &0.05\\
4 & 0.10 & 0.00 & 0.00 (0.00)& 0.00 (0.00)& 0.00 (0.00) & 0.04\\
5 & 0.10 & 0.00 & 0.00 (0.00)& 0.00 (0.00)& 0.00 (0.00) &0.03\\
6 & 0.10 & 0.39 & 0.46 (0.48)& 0.47 (0.52)& 0.50 (0.54) &0.01\\
7 & 0.10 & 0.00 & 0.00 (0.00)& 0.00 (0.00)& 0.00 (0.00) & 0.005\\
8 & 0.10 & 0.01 & 0.01 (0.01)& 0.02 (0.01)& 0.02 (0.01) &0.01\\
9 & 0.10 & 0.04 & 0.04 (0.01)& 0.04 (0.00)& 0.03 (0.00) &0.09\\
10 & 0.10 & 0.01 & 0.00 (0.00)& 0.00 (0.00)& 0.00 (0.00)& 0.005\\
\enddata
\end{deluxetable}

\begin{deluxetable}{cccccc|c}
\tabletypesize{\scriptsize}
\tablecaption{Posterior probabilities for models from Table~\ref{tab:1}; $\Omega_ {\mathrm{m},0} \in \langle 0.25,0.31 \rangle$ }
\tablewidth{0pt}
\tablehead{\colhead{model}&\colhead{prior}&\colhead{posterior SNIa}& \colhead{posterior + CMB}&\colhead{posterior + BAO}&\colhead{posterior + H(z)}&\colhead{posterior SNIa+CMB+BAO+H(z)}}
\startdata
1 & 0.20 & 0.91 & 0.98 (0.88)& 0.99 (0.99)& 0.99 (1.00)&0.84\\
2 & 0.20 & 0.01 & 0.00 (0.00)& 0.00 (0.00)& 0.00 (0.00)&0.02\\
3 & 0.20 & 0.06 & 0.01 (0.11)& 0.00 (0.01)& 0.00 (0.00)&0.06\\
4 & 0.20 & 0.01 & 0.01 (0.00)& 0.01 (0.00)& 0.01 (0.00)&0.04\\
5 & 0.20 & 0.01 & 0.00 (0.00)& 0.00 (0.00)& 0.00 (0.00)&0.04\\
\enddata
\end{deluxetable}

\begin{deluxetable}{cccccc|c}
\tabletypesize{\scriptsize}
\tablecaption{Posterior probabilities for models from Table~\ref{tab:2}; $\Omega_ {\mathrm{m},0} \in \langle 0.25,0.31 \rangle$ }
\tablewidth{0pt}
\tablehead{\colhead{model}&\colhead{prior}&\colhead{posterior SNIa}& \colhead{posterior + CMB}&\colhead{posterior + BAO}&\colhead{posterior + H(z)}&\colhead{posterior SNIa+CMB+BAO+H(z)}}
\startdata
6 & 0.20 & 0.19 & 0.27 (0.26)& 0.35 (0.50)& 0.45 (0.91)&0.07\\
7 & 0.20 & 0.04 & 0.00 (0.00)& 0.00 (0.00)& 0.00 (0.00)&0.03\\
8 & 0.20 & 0.05 & 0.13 (0.04)& 0.24 (0.06)& 0.26 (0.04)&0.13\\
9 & 0.20 & 0.68 & 0.60 (0.70)& 0.41 (0.44)& 0.29 (0.05)&0.74\\
10 & 0.20 & 0.04 & 0.00 (0.00)& 0.00 (0.00)& 0.00 (0.00)&0.03\\
\enddata
\end{deluxetable}

\begin{deluxetable}{cccccc|c}
\tabletypesize{\scriptsize}
\tablecaption{Posterior probabilities for models from Table~\ref{tab:1} and Table~\ref{tab:2}; $\Omega_ {\mathrm{m},0} \in \langle 0.25,0.31 \rangle$ }
\tablewidth{0pt}
\tablehead{\colhead{model}&\colhead{prior}&\colhead{posterior SNIa}& \colhead{posterior + CMB}&\colhead{posterior + BAO}&\colhead{posterior + H(z)}&\colhead{posterior SNIa+CMB+BAO+H(z)}}
\startdata
1 & 0.10 & 0.81 & 0.91 (0.82)& 0.96 (0.96)& 0.96 (0.97)&0.74\\
2 & 0.10 & 0.01 & 0.00 (0.00)& 0.00 (0.00)& 0.00 (0.00)&0.02\\
3 & 0.10 & 0.07 & 0.01 (0.12)& 0.00 (0.01)& 0.00 (0.00)&0.05\\
4 & 0.10 & 0.01 & 0.01 (0.00)& 0.01 (0.00)& 0.01 (0.00)&0.04\\
5 & 0.10 & 0.01 & 0.01 (0.00)& 0.01 (0.00)& 0.01 (0.00)&0.03\\
6 & 0.10 & 0.02 & 0.02 (0.02)& 0.01 (0.02)& 0.01 (0.02)&0.01\\
7 & 0.10 & 0.00 & 0.00 (0.00)& 0.00 (0.00)& 0.00 (0.00)&0.005\\
8 & 0.10 & 0.00 & 0.00 (0.00)& 0.00 (0.00)& 0.00 (0.00)&0.01\\
9 & 0.10 & 0.07 & 0.04 (0.04)& 0.01 (0.01)& 0.01 (0.01)&0.09\\
10 & 0.10 & 0.00 & 0.00 (0.00)& 0.00 (0.00)& 0.00 (0.00)&0.05\\
\enddata
\end{deluxetable}

As one can conclude final results coming from computation including the shrinking parameters interval procedure are the same as final results coming from computation with parameter intervals assumed at the beginning: the best model (in each set) does not change, but the probability of being the best one are greater for second case. 

As we have written before we used the BIC quantity as an approximation to the minus twice logarithm of the evidence for the SNIa data. This approximation gives good results if in the set under consideration is one favoured model. The problem appears when we have two favoured models with nearly the same probabilities. Such a situation is in the set of all models with $\Omega_{m,0} \in <0,1>$. $\Lambda$CDM model and DGP models have nearly the same values of model probabilities which are the greatest ones in considered set. This enforce us to compute the evidence by the numerical integration for these two models (for SNIa data) to check if our previous conclusion is true. In Table 9 we gathered the values of probabilities obtained after computation of the full Bayesian evidence (case 1) as well as such values obtained when the BIC approximation was used (case 2).

\begin{deluxetable}{cccccc}
\tabletypesize{\scriptsize}
\tablecaption{Posterior probabilities for $\Lambda$CDM and DGP models}
\tablewidth{0pt}
\tablehead{\colhead{model}&\colhead{prior}&\colhead{posterior SNIa}& \colhead{posterior + CMB}&\colhead{posterior + BAO}&\colhead{posterior + H(z)}}
\startdata
$\Lambda$CDM (case1) & 0.50 & 0.63 & 0.57 & 0.54 & 0.52\\
DGP (case1) & 0.50 & 0.37 & 0.43 & 0.46& 0.48\\
$\Lambda$CDM (case2) & 0.50 & 0.57 & 0.50 & 0.47 & 0.45\\
DGP (case2) & 0.50 & 0.43 & 0.50 & 0.53& 0.55\\
\enddata
\end{deluxetable}

As one can see the conclusion changed. The $\Lambda$CDM model is better than the DGP model in the light of all data sets used in paper. The BIC approximation is not enough in this cases, gives us wrong answer.

Finally we compare considered sets of models treating all described before data sets as $N=192+1+1+9$ independent data. In this case the likelihood function has the following form
\begin{displaymath}
 L\propto \exp \left[-\frac{1}{2}\left(\sum_{i=1}^{192}\frac{(\mu_{i}^{\mathrm{theor}}-\mu_{i}^{\mathrm{obs}})^{2}}{\sigma_{i}^{2}}+\frac {(R^{\mathrm{theor}}-R^{\mathrm{obs}})^2}{\sigma_{R}^2} +\frac{(A^{\mathrm{theor}}-A^{\mathrm{obs}})^2}{\sigma_{A}^2}+\sum_{i=1}^{9}\frac{\left( H(z_i) -H_i(z_i) \right) ^2}{\sigma_i ^2} \right) \right].
\end{displaymath}
Here we assumed flat prior for model parameters in the range described at the beginning of this section. We used the $BIC$ as an approximation to $-2lnE$. The results for the cases with $\Omega_{\mathrm{m},0} \in <0,1>$ and $\Omega_{\mathrm{m},0} \in <0.25,0.31>$ for all described before sets of models are gathered in Tables 3, 4, 5, 6, 7, 8.

The results in this analysis confirm that the $\Lambda$CDM model is the best one in the set of models with dark energy as well as the best one in the set of all models for both ranges in $\Omega_{\mathrm{m},0}$. The conclusion that the Cardassian model is the best one in the set of models with modified gravity (for both ranges in $\Omega_{\mathrm{m},0}$) is with contrary with previous inference where the DGP was the best one. The reason of this disagreement is related with the different constrains on the $\Omega_{\mathrm{m},0}$ parameter for various data sets for considered model. In Table 10 we presented the results of parameter estimation performed for all data sets independently as well as for all data sets applied simultaneously for the $\Lambda$CDM, DGP and Cardassian models.

\begin{deluxetable}{cccccc}
\tabletypesize{\scriptsize}
\tablecaption{The best fit values of model parameters}
\tablewidth{0pt}
\tablehead{\colhead{model}&\colhead{SNIa}& \colhead{CMB}&\colhead{BAO}&\colhead{H(z)}&\colhead{SNIa+CMB+BAO+H(z)}}
\startdata
$\Lambda$CDM&$\Omega_{\mathrm{m},0}=0.27^{+0.03}_{-0.03}$&$\Omega_{\mathrm{m},0}=0.22^{+0.04}_{-0.03}$ & $\Omega_{\mathrm{m},0}=0.27^{+0.02}_{-0.02}$ & $\Omega_{\mathrm{m},0}=0.31^{+0.1}_{-0.06}$ & $\Omega_{\mathrm{m},0}=0.28^{+0.02}_{-0.02}$\\
& & & & & \\
\hline
& & & & & \\
DGP& $\Omega_{\mathrm{m},0}=0.17^{+0.03}_{-0.02}$ & $\Omega_{\mathrm{m},0}=0.34^{+0.05}_{-0.05}$ &$\Omega_{\mathrm{m},0}=0.31^{+0.03}_{-0.03}$ & $\Omega_{\mathrm{m},0}=0.25^{+0.1}_{-0.06}$& $\Omega_{\mathrm{m},0}=0.28^{+0.02}_{-0.02}$\\
& & & & & \\
\hline
& & & & & \\
Cardassian&$\Omega_{\mathrm{m},0}=0.31^{+0.08}_{-0.11}$ &$\Omega_{\mathrm{m},0}=0.26^{+0.32}_{-0.32}$ & $\Omega_{\mathrm{m},0}=0.27^{+0.39}_{-0.39}$ &$\Omega_{\mathrm{m},0}=0.01^{+0.16}_{-0.16}$ & $\Omega_{\mathrm{m},0}=0.28^{+0.02}_{-0.02}$\\
& & & & & \\
 & $n=-0.15^{+0.31}_{-0.39}$ & $n=0.08^{+1.19}_{-1.19}$ &  $n=0.00^{+3.45}_{-3.45}$&  $n=0.65^{+0.08}_{-0.08}$&  $n=0.04^{+0.08}_{-0.08}$\\
\enddata
\end{deluxetable}

\section{Conclusions}

In this paper we gathered ten models of the accelerating Universe. The five of them explain the accelerated phase of the Universe in the term of dark energy while the other five explain this phenomenon by the modification of the theory of gravity. We used the Bayesian model comparison method to select the best one in the set of models with dark energy, in the set of models with modified theory of gravity as well as the best one of all of them. The selection based on the SNIa, CMB, BAO and observational H(z) data--we treat posterior probabilities obtained in one analysis as prior probabilities in the next one: information coming from the previous analysis allow us to favor one model over another. We used approximation proposed by Schwarz to the minus twice logarithm of evidence in the case with SNIa data, and numerical integration of the likelihood function within an allowed parameter space (we assumed flat prior probabilities for model parameters) in the other cases. We consider separately cases with $\Omega_ {\mathrm{m},0} \in \langle 0,1 \rangle$ and with $\Omega_ {\mathrm{m},0} \in \langle 0.25,0.31 \rangle$. We made a stricter analysis for models with parameters which intervals width exceed one: we evaluated the evidence for these models for different parameters intervals with minimally width equal one, which do not exceed intervals assumed in the analysis with SNIa data and finally chose the best one from them (with the greatest evidence) to the next analysis. We compare such results with the results obtained in calculation where we treat all data sets as $N=192+1+1+9$ independent data and use BIC as an approximation to the minus twice logarithm of the evidence.

We can conclude that for the case with $\Omega_ {\mathrm{m},0} \in \langle 0.25,0.31 \rangle$ as well as for case with  $\Omega_ {\mathrm{m},0} \in \langle 0,1 \rangle$
\begin{itemize}
    \item the $\Lambda$CDM model is the best one from the set of models with dark energy as well as the best one from the set of all models considered in this paper;
    \item the Cardassian model is the best one from the set with models with modified theory of gravity.
\end{itemize}

\acknowledgements
The paper was supported by the Marie Curie Actions Transfer of Knowledge 
project COCOS (contract MTKD-CT-2004-517186). The authors are very grateful to the referee and to Dr W. Godlowski for helpful discussion and suggestions.

\end{document}